\documentclass[aps,prd,amsmath,amssymb,preprintnumbers,twocolumn]{revtex4}
\usepackage[a4paper,left=20mm,right=20mm,top=25mm,bottom=25mm]{geometry}
\usepackage{graphicx}
\usepackage{color}
\usepackage{subfigure}
\usepackage{pxfonts}
\usepackage{fancyhdr}\usepackage{multirow}
\usepackage{tikz}

\usepackage{youngtab}

\newcommand{\beq}{\begin{eqnarray}}% can be used as {equation} or  {eqnarray}
\newcommand{\eeq}{\end{eqnarray}}

\definecolor{rossoCP3}{cmyk}{0,0.88,0.77,0.40}
\newcommand{\symb}{{\color{rossoCP3}{\vardiamondsuit}}}
\newcommand{\symba}{{\color{rossoCP3}{\spadesuit}}}
\begin{document}

%%%%%%%%%%%%%%%%%%%%%%%%%%%%%%%%%%%%%%%%%%%%%%%%%%%%%%%%%%%%%%%%%%%%%%%%%%%%%%%%%%%%%%%%
\title{\Large \color{rossoCP3}  An Ultraviolet Chiral Theory of the Top \\ for \\ The Fundamental Composite (Goldstone) Higgs}
\author{Giacomo Cacciapaglia$^\symba$}
\email{g.cacciapaglia@ipnl.in2p3.fr}
\author{Francesco Sannino$^\symb$}
\email{sannino@cp3.dias.sdu.dk} 
\affiliation{$^\symba$Univ Lyon, Universit\'e Lyon 1, CNRS/IN2P3, IPNL, 4 rue E.Fermi, F-69622 Villeurbanne Cedex, France;}
\affiliation{$^\symb${ CP}$^3${-Origins} and the Danish IAS, University of Southern Denmark, Campusvej 55, DK-5230 Odense M, Denmark.}
%%%%%%%%%%%%%%%%%%%%%%%%%%%%%%%%%%%%%%%%%%%%%%%%%%%%%%%%%%%%%%%%%%%%%%%%%%%%%%%%%%%%%%%%

\begin{abstract}
We introduce a scalar--less anomaly free chiral gauge theory that serves as  natural ultraviolet completion of models of fundamental composite (Goldstone) Higgs dynamics. The new theory is able to generate the top mass and furthermore features a built-in protection mechanism that naturally suppresses the bottom mass. At low energies the theory predicts new fractionally charged fermions, and a number of  four-fermion operators that, besides being relevant for the generation of the top mass, also lead to an intriguing phenomenology for the new states predicted by the theory. 
 \\[2mm]
{\footnotesize\it Preprint: CP$^3$-Origins-2015-032 DNRF90, DIAS-2015-32}
\end{abstract}

\maketitle

The discovery of the Higgs boson in 2012 is a historical turning point, that provided us the first direct handle on the origin of the Brout-Englert-Higgs mechanism. To date, however, the Large Hadron Collider (LHC) experiments have not been able to precisely test this paradigm because of the limited precision attained on the measurement of the Higgs couplings. Furthermore, the Standard Model (SM) is neither a fundamental theory nor able to explain physical observations such as the missing dark mass  and matter--antimatter asymmetry, thus a more fundamental description of nature is highly wanted. 

Models of composite (Goldstone) Higgs dynamics constitute a time-honoured possibility~\cite{Weinberg:1975gm,Susskind:1978ms,Kaplan:1983fs,Kaplan:1983sm}.  However, a satisfactory theory of fermion mass generation is still missing.  
One of the hardest problems to solve, for the composite paradigm, is to generate the observed hierarchy between the heavy top-quark mass and the very light ones for the remaining SM fermions. 

It is therefore tremendously important to construct sufficiently simple, ultraviolet (UV) complete, theories able, in a natural way, to address the top mass generation. This is exactly what we will consider here. In fact, we will construct a unified anomaly-free gauged chiral extension of the SM  featuring fermionic matter that explains the top mass generation and simultaneously leads to the composite (Goldstone) Higgs realisation. We will also discuss further phenomenological implications of the theory. 

We, therefore, start with the setup investigated in \cite{Cacciapaglia:2014uja} according to which the fundamental composite dynamics (FCD) constituting the Higgs sector of the SM is an $SU(2)_{\rm TC} = Sp(2)_{\rm TC}$ gauge theory featuring two new Dirac fermions transforming according to the fundamental representation of the gauge group~\cite{Appelquist:1999dq,Ryttov:2008xe, Galloway:2010bp}. 
Within this simple FCD, the Higgs particle could  naturally emerge as mostly a pseudo-Nambu Goldstone Boson (pNGB) \cite{Kaplan:1983fs,Kaplan:1983sm,Galloway:2010bp},  or as the lightest composite scalar fluctuation of the fermion condensate, like in Technicolor (TC) inspired theories \cite{Appelquist:1999dq,Ryttov:2008xe}. In general, it is a linear combination of both states. In fact, one can show that any underlying four-dimensional composite pNGB nature of the Higgs is always accompanied by a TC-like limit at the fundamental level \cite{Cacciapaglia:2014uja}. They just differ in the dynamical alignment of the electroweak symmetry and its embedding in the larger global symmetry of the fundamental theory. It is possible, however, to construct TC realisations that do not admit a Goldstone Higgs limit. 

The model $SU(2)_{\rm TC}$ constitutes  the minimal realisation of composite pNGB Higgs and TC models  in terms of an underlying fundamental dynamics.  By minimal we mean that it is based on the smallest asymptotically free gauge group with the smallest number of fermions needed to accomplish the required dynamics \cite{Appelquist:1999dq,Ryttov:2008xe, Galloway:2010bp}. Since the representation is pseudo-real, 
the new fermions can be described as 4 Weyl fermions $Q^i$, so that the global symmetry of the fermionic sector is $SU(4)$. The additional classical 
$U(1)$ global symmetry is anomalous at the quantum level. Further interesting physical consequences stem from the  
topological sector \cite{DiVecchia:2013swa}.  Because $SU(2)$ can be viewed as the first of the symplectic groups \cite{Sannino:2009aw}  the 
phenomenological analysis, and model building, can be generalised to $Sp(2N)$~\cite{Barnard:2013zea}.  First principle lattice simulations have confirmed that the $SU(2)$ theory with two Dirac fermions breaks the underlying global symmetry $SU(4)$ to $Sp(4)$ and have further provided crucial insight on the spectrum of spin-one resonances  and Goldstone scattering amplitudes \cite{Arthur:2014lma,Arthur:2014zda,Hietanen:2014xca,Hietanen:2013fya,Lewis:2011zb}.  Note that while the observed Higgs boson is identified with a light meson in the pNGB limit, in the TC limit one needs to rely on the presence of a light bound state with appropriate couplings to the SM particles: the light mass may derive from the presence of a near-conformal behaviour of the theory, broken by the condensate.

We  now extend the model to generate the top mass without yielding, in first approximation, masses for the remaining SM fermions.  In particular we shall concentrate on generating fermion masses via four-fermion interactions bilinear in the elementary top-quark fields, as in Ref.s~\cite{Ryttov:2008xe, Galloway:2010bp,Cacciapaglia:2014uja}.  The masses of the light quarks and leptons can be generated by similar interactions, which are generated at a higher scale. Alternatively, one could consider couplings linear in the quark fields, as done in~\cite{Barnard:2013zea,Ferretti:2013kya}, so that the top acquires its mass via partial compositeness~\cite{Kaplan:1991dc}, but we will not pursue this possibility here. 
 
 Our novel UV composite chiral completion of the top mass abides the following model building conditions: it must 
\begin{itemize}
\item be an anomaly free chiral gauge theory;
\item feature only fermionic matter;
 \item unify ordinary colour and the FCD colour;
\item generate the top mass but no other fermion masses.
 \end{itemize}
 Chiral  gauge theories can dynamically self-break their gauge interactions~\cite{Raby:1979my} (see for example \cite{Appelquist:2000qg}) allowing, at least in principle, the possibility of an underlying theory which is truly free of elementary scalars. This is the reason why we require the theory to be chiral, since it has a fair chance to be a true solution of the SM hierarchy problem. 

\begin{table}[tb]
\begin{center}
\begin{tabular}{c|ccc|c}
 & $SU(2)_{\rm TC}$ & $SU(3)_{\rm c}$ & $SU(2)_L$ & $U(1)_{\rm Y}$ \\
 \hline
 $Q_L = (U_L, D_L)$ & ${\tiny{\yng(1)}}$ & $1$ & ${\tiny{\yng(1)}}$ & $0$ \\
 $U_R^c$ & ${\tiny{\yng(1)}}$ & $1$ & $1$ & $-1/2$ \\
 $D_R^c$ & ${\tiny{\yng(1)}}$ & $1$ & $1$ & $1/2$ \\
 \hline
 $q_L = (t_L, b_L)$ & $1$ & ${\tiny{\yng(1)}}$ & ${\tiny{\yng(1)}}$ & $1/6$ \\
 $t_R^c$ & $1$ & $\overline{\tiny{\yng(1)}}$ & $1$ & $-2/3$ \\
 $b_R^c$ & $1$ & $ \overline{\tiny{\yng(1)}}$ & $1$ & $1/3$ \\
 $l_L =(\nu_L, \tau_L)$ & $1$ & $1$ &  ${\tiny{\yng(1)}}$ & $-1/2$ \\
 $\tau_R^c$ & $1$ & $1$ & $1$ & $1$ \\
 \hline
\end{tabular}
\caption{Fermion content of the low energy theory giving rise to SU(4)/Sp(4) model, including the 3rd generation of SM fermions.} \label{tab:FCD}
\end{center}
\end{table}
We summarise in Table~\ref{tab:FCD} the fermionic matter content that includes the FCD fermions and their charges/transformation properties with respect to the relevant SM gauge interactions and FCD gauge group. We also show the charges/transformation properties of the third family of SM fermions, which will be partly embedded in the UV completion. With this choice of fermions and their transformation properties the overall theory is free from gauge and gravitational anomalies.

We wish now to embed the gauge groups and fermion content of Table~\ref{tab:FCD} into an anomaly free chiral model that naturally generates the crucial 4-fermion interactions responsible to give mass to the top, but forbids the ones responsible for the bottom mass.  
Our strategy is to embed the $SU(2)_{\rm TC}$ and $SU(3)_{\rm c}$, just for the up-type techni--quarks and for the top, in a unified $SU(5)$ group. The gauge group is then extended to be  
\begin{multline} \label{eq:SB}
SU(5)\times SU(2)'_{\rm TC}\times SU(3)'_{\rm c} \times U(1)_{\rm X} \\ \supset SU(2)_{\rm TC} \times SU(3)_{\rm c} \times U(1)_{\rm Y}
\end{multline}
with the weak isospin $SU(2)_{\rm L}$ acting as an external group.
The $SU(2)_{\rm TC}$ group of the FCD theory is thus the diagonal $SU(2)$ of the $SU(2)'_{\rm TC}$ and the $SU(2)$ subgroup of $SU(5)$, and similarly for $SU(3)_{\rm c}$.
The natural assignment is to embed the $SU(2)_L$ doublets $Q_L$--$q_L$, and the singlets $U_R$--$t_R$, in multiplets of SU(5):
\beq
\left( \begin{array}{cc}
U_L & t_L \\
D_L & b_L
\end{array} \right) = \psi_1\,, \quad \left( \begin{array}{cc} U_R^c & t_R^c \end{array} \right) = \psi_2\,.
\eeq
Here $\psi_1$ transforms as a ${\bf 5}$ of SU(5), while $\psi_2$ as a ${\bf \overline 5}$.  We also define $\psi_3 = D_R^c$ to be a singlet of SU(5).  The ordinary hypercharge is defined as the sum of the diagonal generator of SU(5) and the U(1)$_X$ charge, with normalisation:
\beq
Y = 
- \frac{1}{30} \left( \begin{array}{ccccc}
3 & & & & \\
& 3 & & & \\
& & -2 & & \\
& & & -2 & \\
& & & & -2
\end{array} \right) + X\,,
\eeq
with the $X$ charge assignments given in Table~\ref{tab:ETC}.
\begin{table}[tb]
\begin{center}
\begin{tabular}{c|cccc|c}
 & $SU(5)$ & $SU(2)'_{\rm TC}$ & $SU(3)'_{\rm c}$ & $SU(2)_L$ & $U(1)_{\rm X}$ \\
 \hline
 $\psi_1$ & ${\tiny{\yng(1)}}$ & $1$ & $1$ & ${\tiny{\yng(1)}}$ & $1/10$ \\
 $\psi_2$ & $\overline{\tiny{\yng(1)}}$ & $1$ & $1$ & $1$ & $-3/5$ \\
 $\psi_3$ & $1$ &  ${\tiny{\yng(1)}}$ & $1$ & $1$ & $1/2$ \\
 $\psi_4$ &  $\overline{\tiny{\yng(1,1)}}$ & $1$ & $1$ & $1$ & $2/15$ \\
 $\psi_5$ & $1$ & $1$ & $1$ & $1$ & $-1/3$ \\
 $\psi_6$ & $1$ &  ${\tiny{\yng(1)}}$ &  ${\tiny{\yng(1)}}$ & $1$ & $-1/6$ \\
 $\psi_7$ & $1$ & $1$ &  $\overline{\tiny{\yng(1)}}$ & $1$ & $0$ \\
 \hline
 $b_R^c$ & $1$ & $1$ & $ \overline{\tiny{\yng(1)}}$ & $1$ & $1/3$ \\
 $l_L$ & $1$ & $1$ & $1$ &  ${\tiny{\yng(1)}}$ & $-1/2$ \\
 $\tau_R^c$ & $1$ & $1$ & $1$ & $1$ & $1$ \\
 \hline
\end{tabular}
\caption{Fermion content of the UV complete theory.} \label{tab:ETC}
\end{center}
\end{table}

\subsection*{Gauge Anomaly Free Chiral Spectrum}

The theory is not yet gauge anomaly free. We need therefore to add new fermion fields in order to cancel the various gauge anomalies. The complete list is displayed in Table~\ref{tab:ETC}. 
To summarise
\begin{itemize}
\item[-]  {\it SU(5) anomalies:} they arise due to the imbalance between two ${\bf 5}$ and one ${\bf \bar{5}}$ associated respectively to $\psi_1$ and $\psi_2$. One natural possibility is to add a ${\bf \bar{5}}$ with $X$-charge $2/5$, which contains an $SU(2)_{\rm TC}$ doublet with hypercharge $Y = 1/2$ ($\sim D_R^c$), and a QCD anti-triplet with hypercharge $Y = 1/3$ ($\sim b_R^c$). However, because we want to avoid introducing partners of the bottom, we added instead a ${\bf \overline{10}}$ of $SU(5)$ with $X$-charge $2/15$, denoted by $\psi_4$,  which decomposes under $SU(2)_{\rm TC} \times SU(3)_{\rm c} \times U(1)_{\rm Y}$ as:
\beq
\overline{\tiny{\yng(1,1)}} = (1,1,1/3) \oplus (2,\overline{3},1/6) \oplus (1,3,0)\,.
\eeq
We will call these fermions respectively $f_5$, $f_6$ and $f_7$. 
\item[-] the fields $\psi_{5,6,7}$ are  added in order to allow  mass term generations for the components of the $\psi_4$, i.e. the ${\bf \overline{10}}$,  after  symmetry breaking. They contribute to cancelling the anomalies of the other gauge groups, in particular $SU(3)'_{\rm c}$ and $U(1)_{\rm X}$.

\end{itemize}
With the above assignments, all the anomalies, including the gravitational one, vanish nontrivially. It is interesting to note that $\psi_1$, $\psi_2$ and $\psi_4$  make up a generalised Georgi-Glashow SU(5) chiral theory~\cite{Georgi:1974sy} with an extra vector-like fermion in the fundamental representation that is known to be gauge-anomaly free \cite{Bars:1981se}, as summarised in \cite{Appelquist:2000qg} along with possible non-trivial infrared phases.  

We then assume that, at an energy scale $\Lambda_{UV}$, the gauge groups are broken following the pattern in Eq.~\eqref{eq:SB}, leaving behind a number of massive gauge bosons. In particular we are interested in the off-diagonal gauge bosons from the breaking of $SU(5)\to SU(3) \times SU(2)$, which transform like a doublet of $SU(2)_{\rm TC}$ and a triplet of QCD colour (with hypercharge $1/6$). These gauge bosons, that we call $E_\mu$, couple to the following femionic current:
\beq
i \frac{g_5}{\sqrt{2}}\ E_\mu\ J^\mu_E + h.c.\,,
\eeq
where $g_5$ is the SU(5) gauge coupling and
\beq
J^\mu_E = \bar{q}_L \sigma^\mu Q_L - \bar{U}_R^c \sigma^\mu t_R^c + \bar{f}_5 \sigma^\mu f_6 + \bar{f_6} \sigma^\mu f_7\,.
\eeq
Once $E_\mu$ is integrated out, the low energy Lagrangian will contain the following four-fermion interactions
\begin{multline} \label{eq:4fermi}
\mathcal{L}_{\rm 4-fermi} = - \frac{g_5^2}{2 M_E^2}\ J_E^\mu J^\dagger_{E,\mu} = \\ 
= \frac{g_5^2}{2 M_E^2}\ \left( (\bar{q}_L \sigma^\mu Q_L ) ( \bar{t}_R^c \sigma_\mu U_R^c) + h.c. + \dots \right)\,,
\end{multline}
where we singled out the term that, once Fierzed, generates the appropriate mass for the top upon $SU(2)_{\rm TC}$ condensation.
To estimate the scale of $M_E$, we need to estimate the value of the Yukawa coupling of the top: following Naive Dimensional Analysis (see for example \cite{Chivukula:2010tn}) ,
\beq \label{eq:ytNDA}
y_t \sim \frac{g_5^2}{2 M_{E}^2} 4 \pi v^2\,,
\eeq
where $v=2 \sqrt{2} f$ is the chiral symmetry breaking scale (corresponding to the electroweak scale)
of the $SU(2)_{\rm TC}$ driven condensation. More generally, we can identify  $f = \frac{v_{\rm EW}}{2 \sqrt{2} \sin \theta}$, where $v_{\rm EW} = v \sin \theta \sim 246$ GeV is the electroweak scale and $\sin \theta$ is the sine of the angle denoting the alignment of the condensate with respect to the electroweak interactions~\cite{Cacciapaglia:2014uja}. Knowing that $y_t \sim 1$, we obtain:
\beq \label{eq:ME}
\frac{M_E}{g_5} \sim \frac{\sqrt{2 \pi} v_{\rm EW}}{\sin \theta} = \frac{620~\mbox{GeV}}{\sin \theta}\,.
\eeq
The lowest mass can be achieved in the TC limit of the model, for which $\sin \theta = 1$, while in the pNGB Higgs limit the angle is bound to be $\sin \theta \lesssim 0.24$~\cite{Arbey:2015exa} thus giving $M_E \gtrsim g_5\ 2.6$ TeV.
Note that the above estimate would change if the model had a conformal dynamics, which is needed in the TC limit in particular: a large anomalous dimension $\gamma$ of the techni-quark bilinear condensate would add a factor $\left( \frac{\Lambda_{ETC}}{\Lambda_{TC}}\right)^{\gamma} = \left( \frac{M_E}{g_5 4\pi v}\right)^{\gamma}$ to the estimate in Eq.~\ref{eq:ytNDA}, thus allowing to raise the scale where the 4-fermion interactions are generated.

%%%%%%%%%%%
\subsection*{Exotic Vector-like fermions}

Given the gauge anomaly free spectrum described in Table~\ref{tab:ETC}, below the symmetry breaking scale the spectrum contains, besides the fermions in Table~\ref{tab:FCD}, 3 massive vector-like fermions made out of the two Weyl components $f_{5,6,7}$ and $\psi_{5,6,7}$. These latter states are summarised in Table~\ref{tab:VLF} and we use for the overall vector-like states the  $f_{5,6,7}$ nomenclature. 
\begin{table}[h!]
\begin{center}
\begin{tabular}{c|ccc|c}
 & SU(2)$_{\rm TC}$ & SU(3)$_{\rm c}$ & SU(2)$_L$ & U(1)$_{\rm Y}$ \\
 \hline
 $f_5$ & $1$ & $1$ & $1$ & $1/3$ \\
 $f_6$ & ${\tiny{\yng(1)}}$ & $\overline{\tiny{\yng(1)}}$ & $1$ & $1/6$ \\
 $f_7$ & $1$ &  ${\tiny{\yng(1)}}$ & $1$ & $0$ \\
  \hline
\end{tabular}
\caption{Additional vector-like fermions in the FCD theory.} \label{tab:VLF}
\end{center}
\end{table}

Their mass is generated at the symmetry breaking scale $\Lambda_{\rm UV}$, thus they are parametrically of the same order as the mass of the heavy gauge bosons, like $E_\mu$.
Due to their quantum numbers, the lightest of these states is stable, while decays among them are mediated by the massive gauge bosons $E_\mu$ induced by $SU(5)$  breaking. These fermion masses are generated by the symmetry breaking, thus one can safely assume that they will be heavier that the condensation scale of $SU(2)_{\rm TC}$. The associated decays will be mediated by the 4-fermi interactions of Eq.~\eqref{eq:4fermi}.
The mass hierarchies are model dependent and cannot be precisely determined without an explicit model for the symmetry breaking. In the following we will leave the spectrum open.
The fermion $f_6$ plays a crucial role here because it appears in all the couplings of Eq.~\ref{eq:4fermi} that induce decays, and also because it carries FCD charge, thus it will form bound states below the $SU(2)_{\rm TC}$ condensation scale $f$.

\subsubsection*{$f_6$ and its phenomenology}

The fermion $f_6$ carries FCD colour, thus after the condensation of $SU(2)_{\rm TC}$ it will form bound states. We will consider here the case where it is much heavier than the other techni--quarks, thus the lightest bound states will be mesons:
\begin{itemize}
\item[-] $\Phi_q = f_6 Q_L = (\overline{3}, 2, 1/6)$: this composite scalar has the same quantum numbers as a left-handed stop/sbottom, with, however,  the {\it wrong} colour charge;
\item[-] $\Phi_b = f_6 U_R^c = (\overline{3}, 1, -1/3)$: this new composite scalar transforms as a {\it wrong} colour sbottom;
\item[-] $\Phi_t = f_6 D_R^c = (\overline{3}, 1, 2/3)$: this also transforms as a {\it wrong} colour stop.
\end{itemize}
Here the transformation properties are ordered respectively under  colour, weak isospin and hypercharge. 
The decays will be mediated by 4-fermion interactions obtained after integrating out the massive gauge bosons of $SU(5)$: such interactions can, in fact, connect these states to fermions contained in the 5-plets of $SU(5)$. The relevant operators are:
\beq
\mbox{a)}\; (\bar{f}_7 \sigma^\mu f_6) (\bar{q}_L \sigma_\mu Q_L)\,, & & \mbox{b)}\; (\bar{f}_7 \sigma^\mu f_6) (\bar{U}_R^c \sigma_\mu t_R^c)\,, \nonumber \\
\mbox{c)}\; (\bar{f}_5 \sigma^\mu f_6) (\bar{t}_R^c \sigma_\mu U_R^c)\,, & &  \mbox{d)}\; (\bar{f}_5 \sigma^\mu f_6) (\bar{Q}_L \sigma_\mu q_L)\,. \nonumber
\eeq
The 2-body decay modes mediated by the above operators are listed in Table~\ref{tab:decays}. Some of these decays are suppressed by  numerical factors needed to close a line of techni--quarks. Here, $v_{EW}$ stands for the electroweak vacuum expectation value, while $\mu_L$ and $\mu_R$ are the masses of the light techni--quarks, including the dynamical contributions.

\begin{table}[h!]
\begin{center}
\begin{tabular}{c|c|c}
 decay & operator & factor \\
 \hline
$\Phi_q \to f_7 q_L$ & a) & - \\
$\Phi_q^{\rm up} \to f_7 t_R$ & b) & $v_{EW}$ \\
$\Phi_q \to f_5 \bar{q}_L$ & d) & $\mu_L$ \\
\hline
$\Phi_t \to f_7 t_R$ & b) & $\mu_R$ \\
$\Phi_t \to f_5 \bar{b}_L$ & d) & $v_{EW}$ \\
\hline
$\Phi_b \to f_5 \bar{t}_R$ & c) & - \\
$\Phi_b \to f_5 \bar{t}_L$ & d) & $v_{EW}$ \\
\hline   
\end{tabular}
\caption{Decay modes of the coloured $f_6$ mesons: in the third column we list the expected suppression factor, where $\mu_{L/R}$ are the masses of the light techni--quarks, $Q_L$ and $U_R$--$D_R$.} \label{tab:decays}
\end{center}
\end{table}

The decays of $f_5$ and $f_7$ are also be determined by the interactions in Table~\ref{tab:decays}: for instances, transitions between $f_7$ and $f_5$ will always be mediated by a virtual (or real, depending on the masses) $f_6$ meson.

\subsubsection*{Constraints from Cosmology}

This model predicts the existence of stable states with fractional charges or unusual colour assignments. Their masses are expected to range between $1$ TeV and $\mathcal{O} (10)$ TeV (see Eq.~\ref{eq:ME}). Strong constraints on the existence of such states arise from Cosmology, namely from excessive relic abundance. The case of colour-neutral states with only hypercharge ($f_5$) has been recently studied in~\cite{Langacker:2011db}, where the thermal production of such states is considered. The conclusion is that our case, with charge $1/3$, is excluded because of overabundance. The situation is less constrained if the lightest state is coloured: in that case, the relic abundance is much lower~\cite{Wolfram:1978gp}. Strong constraints come from searches of anomalously heavy nuclei on Earth, however these bounds only apply to the case of integer charges (see discussion in Sec. 5.1 of Ref.~\cite{Berger:2008ti}).

The strong constrains we discussed assume that the stable state is thermally produced and that the annihilation is mediated by SM interactions only. In our case, however, additional annihilation channels via the massive gauge bosons are present that may reduce the relic abundance, in particular if a resonant channel is present.
In Ref.~\cite{Berger:2008ti}, an absolute upper limit on the mass of stable charged states is obtained by saturating the unitarity bound for the cross section: the quoted bound is of $\sim 280$ TeV for a scalar, and $\sim 140$ TeV for a fermion, well above the mass scales expected in our model.
Furthermore, in models of inflation with low reheating temperature, the heavy fermions may not be thermally produced at all.
From these considerations, the only reliable direct test of this model will come from the LHC that may be able to produce the heavy fermions and/or the heavy gauge bosons deriving from the $SU(5)$ symmetry breaking. In particular, the phenomenology of the coloured mesons is close to that of stops in supersymmetry, which are bounded to be heavier than about 700 GeV after Run--I.

\subsection*{Conclusions}
We have constructed an anomaly free chiral gauge theory that  naturally extends minimal models of fundamental composite (Goldstone) Higgs dynamics to generate the top mass. The model also features a protection mechanism that naturally suppresses the bottom mass, and predicts the presence of stable heavy vector-like fermions with fractional charges and/or non-standard colour assignment. We have also checked that Flavour Changing Neutral Currents are under control if the flavour mixing is, at most, generated in the up-sector.  We analysed the composite spectrum of the theory and the four-fermion operators that, besides being relevant for the generation of the top mass, lead to an intriguing phenomenology for the new composite states. The masses of the new states, including the new massive gauge bosons, is expected to range between one TeV and $\mathcal{O} (10)$ TeV, thus potentially within reach of LHC Run--II experiments.

\subsection*{Acknowledgements}

G.C. acknowledges partial support from the {\it D\'{e}fiInphyNiTi - projet structurant TLF}, and the Labex-LIO (Lyon Institute of Origins) under grant ANR-10-LABX-66 and FRAMA (FR3127, F\'ed\'eration de Recherche ``Andr\'e Marie Amp\`ere"). The work of F.S. is partially supported by the Danish National Research Foundation grant DNRF:90.

\end{document}